\PassOptionsToPackage{table}{xcolor}
\documentclass[sigconf,authorversion,nonacm,screen]{acmart}

\renewcommand\footnotetextcopyrightpermission[1]{} % removes footnote with conference information in first column
\AtBeginDocument{%
  \providecommand\BibTeX{{%
    \normalfont B\kern-0.5em{\scshape i\kern-0.25em b}\kern-0.8em\TeX}}}

\usepackage{multirow}
\usepackage{enumitem}
\usepackage{tabularx}
\usepackage{float}

\definecolor{lightgray}{gray}{0.9} % Define the light gray color

\begin{document}

\title{An AI System Evaluation Framework for Advancing AI Safety: Terminology, Taxonomy, Lifecycle Mapping}

\author{Boming Xia}
\orcid{0009-0003-7385-4023}
\affiliation{%
  \institution{CSIRO's Data61 \& University of New South Wales}
  \city{Sydney}
  \country{Australia}
}
% \email{boming.xia@data61.csiro.au}

\author{Qinghua Lu}
\orcid{0000-0002-9466-1672}
\affiliation{%
  \institution{CSIRO's Data61 \& University of New South Wales}
  \city{Sydney}
  \country{Australia}
}
% \email{qinghua.lu@data61.csiro.au}

\author{Liming Zhu}
\orcid{0000-0001-5839-3765}
\affiliation{%
  \institution{CSIRO's Data61 \& University of New South Wales}
  \city{Sydney}
  \country{Australia}
}
% \email{liming.zhu@data61.csiro.au}

\author{Zhenchang Xing}
\orcid{0000-0001-7663-1421}
\affiliation{%
  \institution{CSIRO's Data61 \& Australian National University}
  \city{Canberra}
  \country{Australia}
}
% \email{zhenchang.xing@data61.csiro.au}

\renewcommand{\shortauthors}{B. Xia, et al.}

\begin{abstract}
  The advent of advanced AI underscores the urgent need for comprehensive safety evaluations, necessitating collaboration across communities (i.e., AI, software engineering, and governance).
  However, divergent practices and terminologies across these communities, combined with the complexity of AI systems—of which models are only a part—and environmental affordances (e.g., access to tools), obstruct effective communication and comprehensive evaluation.
  This paper proposes a framework for AI system evaluation comprising three components: 1) harmonised terminology to facilitate communication across communities involved in AI safety evaluation; 2) a taxonomy identifying essential elements for AI system evaluation; 3) a mapping between AI lifecycle, stakeholders, and requisite evaluations for accountable AI supply chain. This framework catalyses a deeper discourse on AI system evaluation beyond model-centric approaches.

\end{abstract}

\keywords{Responsible AI, AI Safety, Evaluation, Benchmarking, AI Testing}

\maketitle

\section{Introduction}
As AI evolves into more Advanced forms, including highly capable General (Purpose) AI and highly capable Narrow AI \cite{AISafetyInstitute2024}, their increasing presence in daily life magnifies safety concerns \cite{bengio2023managing, 2023responsible, lu2023responsible}, highlighting the need for comprehensive safety evaluations.
Such necessity is further echoed in key policy discussions, such as the \textit{US Executive Order on the Safe, Secure, and Trustworthy Development and Use of Artificial Intelligence}~\cite{EO_AISafeSecureTrustworthy2023} and the \textit{Bletchley Declaration}~\cite{BletchleyDeclaration2023} on AI safety, reflecting a global consensus on this matter.

Despite the recognised need for rigorous safety evaluations, challenges persist.
Divergent interpretations and usage of key terms such as ``evaluation'', ``testing'', and ``assessment'' across AI, software engineering (SE), and governance communities hinder a unified approach.
The AI community focuses on model alignment and evaluation, while the SE community emphasises system quality assurance. Meanwhile, the governance community concerns about assessing risks and impacts on people and society.

Moreover, existing evaluation methods and practices are fragmented. On one hand, the prevailing focus on model-level evaluation (e.g., \cite{wang2023decodingtrust, zhou2023don, manakul2023selfcheckgpt, sun2024trustllm}) does not fully capture the complexity of AI systems (i.e., AI-based software systems), which incorporate AI (e.g., AI model(s)) and non-AI components~\cite{compound-ai-blog, EvaluatingLLMSystems2024Blog}.
For example, evaluating object recognition model for parking does not alone ensure the safety of autonomous vehicles, which also needs precise manoeuvring and obstacle avoidance.
On the other hand, the internal evaluation conducted by AI Producers tends to exclude other stakeholders (e.g., AI Deployer) within the AI supply chain.
This exclusion can lead to a lack of transparency and reduced accountability, potentially resulting in oversight of critical safety and performance issues that could affect end-users.

Further complicating these issues is the inclusion of environmental affordance factors in AI systems, such as external tool access and safety guardrails. These context-specific factors extend beyond model and challenge consistent evaluation across various deployments \cite{sharkey2024causal}.
Consequently, evaluation needs to consider the unique environmental and operational contexts, reflecting the specific requirements and expectations of its intended uses.

To bridge the gaps, this paper highlights the need for a comprehensive AI system evaluation shift, presenting a framework with three key components:
% \begin{itemize}[leftmargin=*, itemsep=1pt, topsep=2pt, after=\vspace{-\baselineskip}]
\begin{itemize}[leftmargin=*]
    \item Harmonising terminology related to AI system evaluation across disciplines for consistent cross-disciplinary communication.
    \item Presenting a taxonomy for comprehensive AI system evaluation at both component and system levels.
    % incorporating a ``role and responsibility'' perspective within the AI supply chain.
    \item Mapping the AI system lifecycle to requisite evaluations, emphasising stakeholder accountability across the AI supply chain.

\end{itemize}

\begin{table*}
\caption{Terms and Definitions Related to AI System Evaluation}
% \vspace{-7pt}
\label{tab:ai-terms-full}
\centering
% {\small
\rowcolors{2}{white}{lightgray}
% \begin{tabularx}{\textwidth}{>{\hsize=0.35\hsize}X>{\hsize=1.65\hsize}X}
\begin{tabularx}{\textwidth}{>{\hsize=0.33\hsize}X>{\hsize=1.67\hsize}X}
\toprule
\textbf{Term} & \textbf{Definition} \\
\midrule
\textbf{Narrow AI} & A type of AI designed and optimised to perform a \textbf{specific task or set of tasks} within a narrow problem domain. \\
\textbf{General AI} & A type of AI that can handle a\textbf{ wide array of tasks and uses}, both intended and unintended by developers. \\
\textbf{Evaluation} & The process of assessing against specific criteria with or without executing the artefacts, including model/system evaluation, capability evaluation, benchmarking, testing, verification, validation, as well as risk/impact assessment.\\
\textbf{Model evaluation} & The process of assessing an AI model against \textbf{predefined specific criteria or general benchmarks} (beyond accuracy), including model capability evaluation, benchmarking, testing, verification, and validation.\\
\textbf{System evaluation} & The process of assessing an AI system against \textbf{predefined specific criteria or general benchmarks} (beyond functional accuracy/correctness), including system capability evaluation, benchmarking, testing, verification, validation, and risk/impact assessment.\\
\textbf{Capability evaluation} & The process of comprehensively assessing a General AI model/system's overall capabilities, including planned, unplanned, emerging, or dangerous capabilities (\textbf{beyond functions}).\\
\textbf{Benchmarking} & The process of conducting a type of general evaluation that comparatively assesses an AI model/system’s performance against a set of \textbf{predefined standards or reference tasks} as a performance measure. \textbf{Benchmarks} are public datasets and metrics that specify tasks and objectives for AI systems, serving as a standard point of comparison.\\
\textbf{Testing} & The process of \textbf{executing} an AI model/system to verify and validate that it exhibits expected behaviours across a set of appropriately selected test cases. These test cases can be under normal conditions or stress and adversarial conditions such as via red teaming.
A \textbf{test case} is the specification of all essential entities for testing: inputs, testing procedures, and the expected outcomes. A collection of test cases forms a \textbf{test suite}.\\
\textbf{Verification} & The process of confirming AI models/systems meet \textbf{specified requirements}, including dynamic (execution-based) and static (non-execution) verification.\\
\textbf{Validation} & The process of confirming that AI models/systems meet intended uses/expectations \textbf{by its users}.\\
\textbf{Risk assessment} & The systematic process of identifying and evaluating the likelihood and potential consequences of events or actions within AI systems that could lead to harm.\\
\textbf{Impact assessment} & The systematic process for identifying and evaluating the wider and longer term effects that AI systems may have on individuals, communities and society across economic, social, and environmental dimensions.\\
% \end{itemize} \\
\bottomrule
\end{tabularx}
% }
% \vspace{-10pt}
\end{table*}

\section{Harmonised Terminology}
\label{Sec:Term}
Collaborative efforts across multiple disciplines are essential for ensuring AI safety through evaluation. To address the variability in terminology across fields, we have harmonized the terms in Table \ref{tab:ai-terms-full}, drawing from diverse sources (e.g., \cite{isoiec22989:2022, NIST_AIRMF, IEEE2014SWEBOK}).

In AI, ``model validation'' assesses initially trained model using a validation dataset, ``model testing'' examines generalisability on a separate test dataset post-training, and ``model evaluation'' broadly assesses a model's performance/accuracy using various metrics and methodologies.

Extending to a holistic and system-level context, these terms take on broader meanings (see Table \ref{tab:ai-terms-full}).
``Evaluation'' becomes a comprehensive process covering different evaluation strategies, encompassing both static (without execution) and dynamic (with execution) dimensions.
Moreover, the evaluation for AI model/system extends beyond mere statistical accuracy to cover functional \textbf{accuracy/correctness},  \textbf{quality attributes and risks}, and broader \textbf{capabilities} beyond functional tasks.

``\textbf{Accuracy}'' is reconceived as \textbf{task completion fidelity}, capturing not only statistical precision but also how well an AI model/system achieve its tasks.
``\textbf{Correctness}'' emphasises \textbf{system-level operational integrity}, pertaining to the system fulfilling its specified functionalities and user expectations.
Other than these functional aspects, \textbf{quality attributes} (e.g., reliability, robustness, security), as non-functional requirements, detail the overall quality and effectiveness of an AI system against predefined criteria, emphasising immediate outcomes while also considering long(er)-term adaptability, maintainability, and sustainability.
Conversely, \textbf{risk} assessment is concerned with the potential immediate harms, while impact assessment emphasises the longer-term and broader effects.
Meanwhile, \textbf{capability} evaluation extends to both designed and emergent functionalities, including potentially dangerous ones, by examining an AI's adaptability and evolution beyond initial training.

Benchmarking, a crucial evaluation type, is especially pivotal for evaluating General AI which spans various tasks. This breadth necessitates a comprehensive examination of aspects such as quality/risk (e.g., fairness and bias), accuracy/correctness (e.g., a translation AI's fidelity and the translation software's operational integrity), and (intended and unintended) capabilities \cite{weidinger2023sociotechnical}. By systematically comparing against diverse baselines and competitors, benchmarking ensures not only performance excellence but also ethical integrity across different contexts (e.g., scenarios, use cases). This \textbf{context-based evaluation} is essential for obtaining meaningful and actionable insights, ensuring the relevance and applicability of the benchmarking results to real-world deployments.

\section{AI System Evaluation: A Taxonomy}
\begin{figure*}
    \centering
    \includegraphics[width=\linewidth]{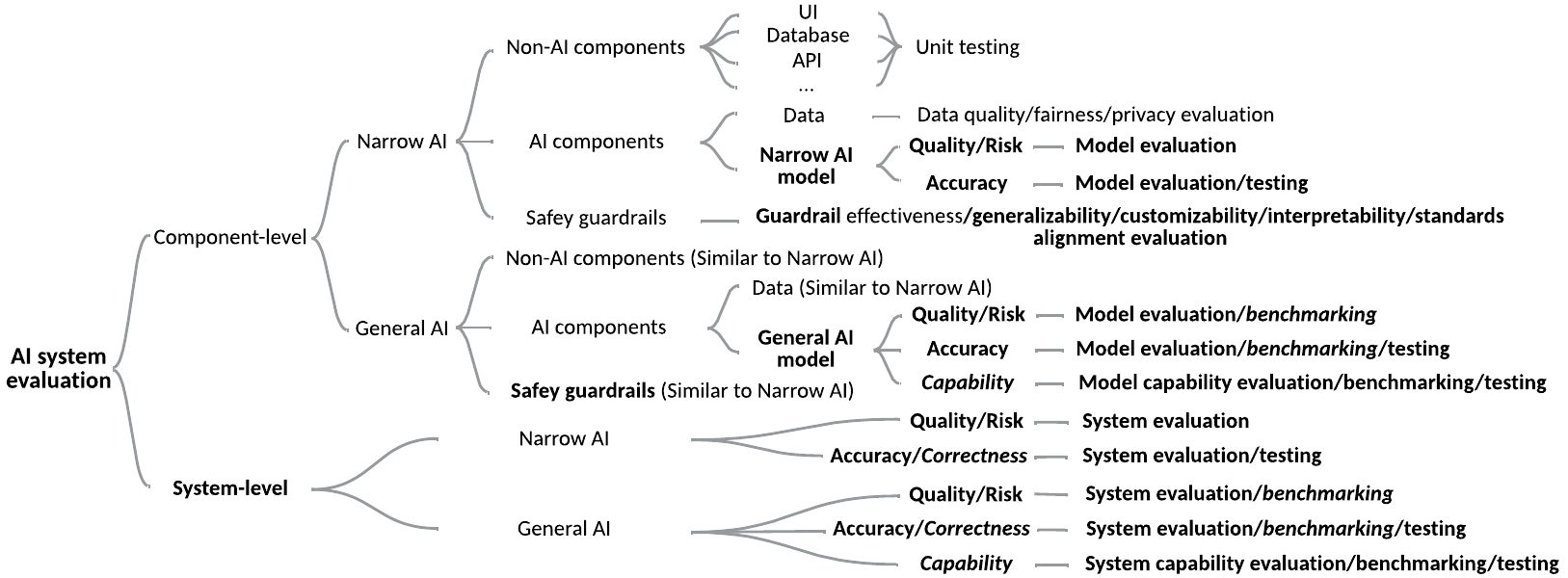}
    \caption{A Taxonomy for AI System Evaluation}
    \label{fig:taxo}
\end{figure*}

Fig. \ref{fig:taxo} presents a taxonomy for evaluating AI systems at component and system levels, emphasising \textbf{key areas}.

\subsection{Component-Level Evaluation}
Evaluation at this level includes both AI and non-AI components.

\subsubsection{\textbf{Non-AI Components Evaluation}}

Non-AI components, such as user interfaces and application programming interfaces, are crucial for the AI system's integrity and necessitate unit tests akin to traditional software. As our taxonomy emphasises a model versus system perspective, the integration of AI and non-AI components, although critical, falls outside this paper's scope.

\subsubsection{\textbf{AI Components Evaluation - Data}}
% This subsection outlines the evaluation for training data.
We consider data as an AI component because it directly influences model training, performance, and behaviour.
Data evaluation mainly addresses general \textbf{data quality}, \textbf{fairness}, and \textbf{privacy}. General quality evaluation involves verifying accuracy (error-free reflection of real-world phenomena), completeness (coverage of necessary features for training), consistency (uniform standards across the dataset), and timeliness (up-to-dateness relevant to the application).
Data bias evaluation includes selecting and applying fairness metrics (e.g., group and individual) to evaluate equity across demographics \cite{chen2023fairness}.
Evaluating data privacy requires examining of how data is collected, stored, processed, and shared, ensuring compliance with standards and/or regulatory requirements (e.g., EU GDPR \cite{eu2016gdpr}).

\subsubsection{\textbf{AI Components Evaluation - Model}}
We focus on evaluating Narrow and General AI models, while noting the relevance of other AI components like ML programs and libraries \cite{chen2023fairness}.

\textit{3.1.3.1   \textbf{Narrow AI models}} are evaluated on \textbf{Quality/Risk} and \textbf{Accuracy}, given their specialised functions in defined domains.

\textbf{Quality\slash Risk - Model Evaluation}. 
Quality evaluation scrutinises a model’s inherent properties, assessing attributes like robustness, security, and fairness, aligning with model specifications.
Risk evaluation probes into how deficiencies in these attributes could lead to negative outcomes (e.g., fairness vs. bias). This ensures the model positively behaves on both technical and social dimensions.

\begin{figure*}
    \centering
    \includegraphics[width=0.83\linewidth]{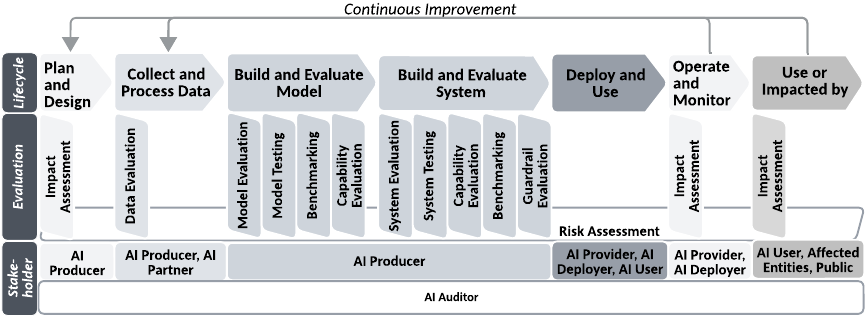}
    \caption{AI System Evaluation (Supply Chain Perspective)}
    \label{fig:supplychain}
\end{figure*}

\textbf{Accuracy - Model Evaluation: Testing}.
Model accuracy evaluation complements non-functional Quality/Risk evaluation. This process hinges on testing which systematically measures a model’s functional accuracy against predefined expectations.
Evaluating model accuracy necessitates testing across selected test cases (i.e., a test suite), incorporating datasets that reflect real-world complexities and metrics for quantifying accuracy. 
Metrics range from general (precision, recall, F1 score) to domain-specific (e.g., BLEU for natural language processing \cite{papineni2002bleu}), tailored to the model's purpose. This ensures a reproducible, context-aware evaluation, aligning model performance with practical applicability.

\textit{3.1.3.2   \textbf{General AI models}}, designed for versatile tasks, require evaluation covering \textbf{Quality/Risk}, \textbf{Accuracy}, and \textbf{Capability}.

\textbf{Quality/Risk - Model Evaluation: Benchmarking}.
Benchmarking distinguishes General AI's evaluation by employing standardised criteria and metrics tailored to its versatile nature, contrasting with Narrow AI's focused scope. This approach ensures General AI models excel in adaptability, transparency, and interoperability—attributes critical for operation across varied contexts—while also meticulously assessing complex risks like copyright infringement due to their extensive training data \cite{zhang2023navigating}. Through comparative analysis against established benchmarks, this method highlights General AI's broader quality and risk spectrum, ensuring these models meet stringent technical and ethical benchmarks beyond the narrower confines of Narrow AI models.

\textbf{Accuracy - Model Evaluation: Benchmarking \& Testing}. The accuracy evaluation of General AI models, like Large Language Models (LLMs), starts with benchmarking against standardised benchmarks (e.g., GLUE \cite{wang2018glue} for linguistic abilities) to establish a broad accuracy baseline and compare capabilities with rival systems. This initial step sets expectations for the model’s accuracy across a diverse array of tasks, highlighting areas for improvement. Subsequent testing, utilising a tailored selection of datasets and metrics, probes the model’s adaptability and output quality in both typical and edge-case scenarios, providing a comprehensive view of its accuracy.
Combining benchmarking and testing provides a thorough understanding of model accuracy, differentiating General AI's broad-spectrum evaluation from Narrow AI's.

\textbf{Capability - Model Capability Evaluation: Benchmarking \& Testing}.
Capability evaluation transcends mere functionality, spotlighting the broader capabilities of General AI, such as reasoning, learning, and ethical decision-making.
Similarly, benchmarking initially offers a baseline quantification of such broad capabilities against selected standards, providing a preliminary understanding of the model's abilities. Subsequent red teaming testing challenges the AI with diverse scenarios to identify any unintended, hazardous behaviours or ethical vulnerabilities. This two-step process ensures a thorough capability evaluation, highlighting areas for improvement and ethical considerations prior to deployment.
% This process not only highlights the model’s designed capabilities but also help uncover emergent ones, guiding stakeholders in deployment decisions.

\subsubsection{\textbf{Safety Guardrails Evaluation}}
Amid growing AI safety concerns, safety guardrails and their evaluation become critical in advanced AI systems, particularly in LLMs \cite{vidgen2024introducing}. These mechanisms, whether AI-driven or otherwise, play a crucial role in maintaining safety and ensuring ethical standards. Our focus shifts from the effectiveness of guardrails when applied to AI models/systems to their \textbf{inherent characteristics}. The discourse also highlights the importance of system-level (outside-model) guardrails, external mechanisms set to define operational boundaries and ensure safety compliance, over (in-)model guardrails integrated during training.

The evaluation addresses four critical attributes: \textbf{Generalisability} across different AI models/systems and contexts, ensuring they remain effective under varied operational scenarios; \textbf{Customisability}, which allows guardrails to be ``fine-tuned'' to the specific needs of each deployment \cite{zhang2024shieldlm}. For example, autonomous driving guardrails might issue a warning in a certain region when hands are off the wheel but initiate slowing down in another, reflecting deployment contexts; \textbf{Interpretability}, which determines whether human users can understand the guardrails. This is crucial for ensuring that users can comprehend the safety measures in place and trust the system's operations. Clear and understandable guardrails enhance user trust and facilitate better interaction with the AI system; and \textbf{Standards alignment}, ensuring that guardrails adhere to established safety, ethical, and regulatory frameworks, making them not just technically sound but also ethically responsible and compliant with legal requirements.

\subsection{System-Level Evaluation}
System-level evaluation requires analysing AI systems' complexity, including environmental affordances and components' interplay.

\subsubsection{\textbf{Narrow AI System Evaluation}}
Similar to Narrow AI models, evaluation of these systems excludes capability considerations.

\textbf{Quality/Risk - System Evaluation}:
Narrow AI system evaluations are designed to ensure that, as cohesive units comprising various components, these systems adhere to high quality standards and effectively address associated risks. This approach broadens the scope beyond individual model to incorporate additional critical operational attributes such as usability, interoperability, and maintainability. It emphasises the systemic qualities essential for overall system effectiveness, ensuring the systems are robust, user-friendly, and seamlessly integrated into their intended environments.

\textbf{Accuracy/Correctness - System Evaluation: Testing}.
This evaluation focuses on assessing the system's functional correctness and accuracy through rigorous tests that compare outcomes against expected results. It extends beyond individual model tests by encompassing a series of system-wide test cases under varied conditions to evaluate the system's functionality and accuracy, ensuring the system fulfils user expectations and functions correctly.

\subsubsection{\textbf{General AI System Evaluation}}
General AI systems require comprehensive evaluation beyond Narrow AI's targeted focus.

\textbf{Quality/Risk - System Evaluation: Benchmarking}.
This benchmarking process transcends the task-specific evaluations of Narrow AI and the isolated scrutiny of individual General AI models. 
It employs a systematic approach to compare these systems against recognised benchmarks (e.g., EU AI Act, ISO/IEC 25010:2023), focusing on quality attributes significant for General AI's broad application range—such as reliability and security—as well as adherence to ethical principles like privacy, and transparency.

\textbf{Accuracy\slash Correctness - System Evaluation: Testing\slash\linebreak[0] Benchmarking}.
The evaluation of General AI systems combines benchmarking with general criteria and specific testing, akin to approaches used for General AI models, but with a focus on system functional correctness in addition to model-level accuracy. Benchmarking assesses adherence to wide-ranging standards, while targeted testing scrutinises the system's operational integrity across selected scenarios, ensuring both overall functionality and task-specific fidelity.

\textbf{Capability - System Capability Evaluation: Benchmarking\slash Testing}: 
Capability evaluation for General AI systems assesses how environmental affordance factors—like guardrails, access to tools, and user interactions—affect overall system capabilities compared to those of individual models. Similar to General AI models, benchmarking and testing (i.e., red teaming) examine both intended and unintended/emergent system capabilities for safe deployment.

\subsection{Mapping Evaluations to Lifecycles and Stakeholders}
The taxonomy introduced reveals a critical gap: the nuanced interplay between various evaluation dimensions—model\slash system level and general\slash context-specific—is not fully addressed. This indicates existing evaluation practices might not fully grasp the complexity of AI systems, underscoring the need for an integrated evaluation framework that accommodates these varied dimensions.

Toward realising this framework, we map the requisite evaluations to AI system development stages and various stakeholders, adapting NIST's AI Risk Management Framework \cite{NIST_AIRMF} for a system-level perspective. This analysis considers the roles and responsibilities of \textbf{organisation-level stakeholders} across the AI supply chain \cite{xia2024responsible}, as detailed in Fig. \ref{fig:supplychain}, illustrating the need for evaluations that span the entirety of the development lifecycle and engage all relevant stakeholders. These stakeholders include:
\textbf{AI Producer}: An entity engaged in the design, development, testing, and supply of AI technologies, including models and components \cite{isoiec22989:2022};
\textbf{AI Provider}: An entity that offers AI-driven products or services, including both platform providers and those offering specific AI-based products or services \cite{isoiec22989:2022};
\textbf{AI Partner}: An entity offering AI-related services, such as system integration, data provisioning, evaluation, and auditing \cite{isoiec22989:2022};
\textbf{AI Deployer}: An \textbf{organisation} that utilised an AI system by making the system or its outputs (e.g., decisions/predictions/recommendations) available to internal or external users (e.g. customers);
\textbf{AI User}: An entity utilising or relying on an AI system, ranging from organisations (e.g., businesses, governments, non-profits) to individuals or other systems. In some contexts, an AI organisation user is equivalent to an AI deployer;
\textbf{Affected Entity}: An entity impacted by the decisions or behaviours of an AI system, including organisations, individuals, communities, and other systems.

This organisation-level stakeholder-focused analysis identifies enhancement areas and lays the groundwork for a integrated evaluation framework. While developing such a framework is beyond this paper's scope, it's essential for promoting safer and more accountable AI development, deployment, and use.

\section{conclusion}
This paper highlights a paradigm shift towards system-level evaluations for AI safety, articulating a robust framework characterised by harmonised terminology, an evaluation taxonomy, and lifecycle mapping. Central to this framework is the recognition that AI models, while crucial, constitute only part of the broader AI system.
This understanding necessitates a more comprehensive evaluative approach that considers the intricate dynamics of AI systems and their extensive supply chains. By focusing on learning-based advanced AI systems, we lay the groundwork for an integrated evaluation framework that addresses the multifaceted complexities of these systems. This paper not only deepens the discourse on AI system evaluation but also aligns with the pressing needs for AI safety.

\bibliographystyle{ACM-Reference-Format}
\bibliography{ref}

\end{document}